\documentclass[usenatbib]{mn2e} 
 \usepackage{color}
 \usepackage{float} 
 \usepackage[fleqn]{amsmath}
 \usepackage{epsfig,floatflt}
 \usepackage{natbib}
 \usepackage{subfigure}
 \usepackage{amssymb}
 \usepackage{multirow}
 \usepackage{ulem}
 \usepackage{bm}
 \usepackage{url}
 \usepackage{placeins}

 \newcommand{\kms}{{km~s$^{-1}$}}

\title[The Environments of local LIRGs]
{The Environments of Local Luminous Infrared Galaxies: Star Formation Rates Increase with Density}

\author[Abiy G. Tekola et al.]
{Abiy G. Tekola$^{1,2,3}$\thanks{abiy@saao.ac.za}, Petri V\"{a}is\"{a}nen$^2$\thanks{petri@saao.ac.za}, Andreas Berlind$^3$\thanks{a.berlind@vanderbilt.edu} \\
$^1$Astronomy Department, and Astrophysics, Cosmology and Gravity Centre (ACGC), University of Cape Town, Private Bag X3,\\
 Rondebosch 7701, South Africa\\
$^2$South African Astronomical Observatory, PO Box 9, 7935 Observatory, Cape Town, South Africa\\
$^3$Department of Physics and astronomy, Vanderbilt University, Nashville, 1807 Station B, Nashville, TN 37235, USA}

\begin{document}
 \maketitle
 \begin{abstract}
 This work studies the environments and star formation relationships of local luminous infrared galaxies (LIRG) in comparison to other types of local and distant ($z\sim1$) galaxies. The infrared (IR) galaxies are drawn from the IRAS sample. The density of the environment is quantified using 6dF and Point Source Catalogue redshift survey (PSCz) galaxies in a cylinder of $2h^{-1}$ Mpc radius and $10h^{-1}$ Mpc length. Our most important result shows the existence of a dramatic density difference between local LIRGs and local non-LIRG IR galaxies. LIRGs live in denser environments than non-LIRG IR galaxies implying that $L_{IR}=10^{11}h^{-2}L_\odot$ marks an important transition point among IR-selected local galaxies. We also find that there is a strong correlation between the densities around LIRGs and their $L_{IR}$ luminosity, while the IR-activity of non-LIRG IR galaxies does not show any dependence on environment. This trend is independent of mass-bin selection. The SF-density trend in local LIRGs is similar to that found in some studies of blue cloud galaxies at $z\sim1$ which show a correlation between star formation and local density (the reversal of the relation seen for local galaxies). This, together with the rapid decline of the number count of LIRGs since $z\sim1$, could mean that local LIRGs are survivors of whatever process transformed blue cloud galaxies at $z\sim1$ to the present day or local LIRGs came into existence by similar process than high redshift LIRGs but at later stage.

\end{abstract}

 \section{Introduction}
It is known that the universal star formation rate (SFR) has been declining since $z\sim1$ \citep{lilly96, madau98, Bell05, Perez05}. In support of this decline, one can mention that highly luminous infrared galaxies (LIRGs; defined as galaxies with infrared luminosities of $(10^{11}h^{-2}L_\odot \leq L_{IR}\leq10^{12}h^{-2}L_\odot)$)\footnote{$L_{IR}=L_{(8-1000) \mu m}$}, which are thought to be the dominant source of the cosmic star formation (SF), are  much more common in the high redshift universe $(z\sim1)$ than in the present Universe \citep{Pozzi04, Perez05}. The decline of the rate of major mergers and minor interactions  since $z\sim1$ has been suggested as the main driver for the decline of SF \citep{Bell05,Bridge07}.
 
 However, recent studies have found that the average merger rate has been constant since $z\sim1$ and merging has little or no effect on the declining SFR \citep{Bundy04,Elbaz07}. There is evidence for the existence of a high fraction of luminous IR  sources that are  not morphologically disturbed at $z\geq0.7$ \citep{Bell05,Elbaz07}. \cite{Bell05} claim that at this redshift more than half of the strongly star forming galaxies are  normal spirals and only 30\%  are interacting systems. Further evidence showed that more than half of the LIRG population at  $z\geq0.5$ are normal spirals \citep{Mell05}. Following these results, the exhaustion of gas was proposed as an alternative explanation for the drop in SFR \citep{Noeske07}. Since the morphology of galaxies is related to their environments, it is important to explore the SF-environment connection of galaxies that dominate the global SF both at high and low redshift (such as LIRGs) to isolate the underlying reason for the drop in the global SFR.  
 
Addressing the difference in SFR in the high and low redshift Universe is compounded by the difference in SFR-density relationships at the two redshifts. SFR increases with environment richness in the high redshift Universe \citep{Elbaz07,coop08,tran2010,li11} in contrast to the case in the low redshift Universe \citep{gomez03, Kauf04}, though other studies have also found  that the SFR-density correlations disappear at high-$z$ \citep[e.g.][]{finn05,cucci06,patel09,tasca09,ferug10}. The explanation of \cite{Elbaz07} for the reversal of SFR-density relation they found is the process of large scale structure formation at the time, while the majority of SF in the present day Universe is not driven by the process of cluster formation. Many authors on the other hand argue that these differences between the high and low redshift SFR-environment trends are the influence of mass down-sizing, and mass-dependent selection effects operating together;  see e.g.\  \citet[][]{sob2010,coop10,peng10} and references therein, for recent discussions. At high redshift, due to higher luminosity detection limits, surveys are biased toward probing a higher mass regime. On the other hand, at low redshift one can easily probe a wider mass range. When we add the effect of mass down-sizing to this, the contradicting results may not be surprising. As a result, a better detection limit, especially at high redshift, is required in order to better understand these processes. 

LIRGs, being the major contributors to the SF budget and being high in number at $z\sim1$  \citep{Elb02}, undoubtedly play  a prominent role in any reversal of the star formation-density relation at high redshift. Understanding the drop of global SFR is not only a matter of understanding SF in the high redshift Universe but also a matter of understanding SF in the local Universe and knowing the exact reason and implication of the rarity of local LIRGs.  Understanding the origin of the  space density differences between local and high redshift LIRGs and even their formation mechanisms at the two redshifts could also give important insight to the origin of the drop of SF in the present Universe.  Unfortunately the environments of either local or distant LIRGs have not been well explored.
 
At the high and intermediate redshift Universe, different studies claim different results regarding the environment of LIRGs. \cite{Marcill05} carried out one of the few extensive studies of the local environments of intermediate-redshift ($0.7\leq z \leq1$) LIRGs and ULIRGs (ultra luminous IR galaxies, having even greater infrared luminosities of $(L_{IR}\geq10^{12}h^{-2}L_\odot)$). They found that at  $z\sim1$, there is  no correlation between LIRG environment and their IR  activity and implied that once  LIRGs are formed their environment does not play a significant role in affecting their star formation properties.  \cite{caputi09a} in turn found that their sample of LIRGs at redshifts $0.6 < z < 0.8$ lived in overdense and more passive environments, while ULIRGs were at underdense and more active environments, compared to other galaxies with similar masses.  \cite{finn2010} meanwhile found that the fraction of LIRGs is higher in fields than in clusters, at a redshift range of  $0.4 < z < 0.8$.

Neither is the environment of local (U)LIRGs well established.  \cite{Tacco02} showed that local LIRGs and ULIRGs are not found in clusters. Correlating a small sample of  LIRGs and  ULIRGs with NED galaxy clusters and groups, they found that LIRGs and ULIRGs are found only in the fields or in small groups of galaxies (less than  ten  members), which is a similar  environment as disky elliptical galaxies. This could imply evolutionary connections between disky  ellipticals and LIRGs/ULIRGs. Another independent study with detailed quantitative analysis supported the idea that most ULIRGs live in environments similar to field galaxies \citep{Zaud07}. However, the same study found some LIRGs and ULIRGs in clusters of Abell richness 0 and 1. \cite{koulou06} found that the brightest IR galaxies have a higher fraction of neighbors within 1~Mpc than a control sample while \cite{hwang2010} in contrast, found that the $L_{IR}$ values of local IR galaxies have no correlation with their environment. 

Generally speaking, the star formation rate of galaxies is a function of two important parameters: environment and gas fraction. For example, SFR (or $L_{IR}$ output) increases when one moves from isolated galaxies with quiescent star formation phase to galaxies with high fraction of gas and merger dominated star formation phase \citep{joha09}. The morphology-environment relation in local LIRGs is consistent with this simple scenario. Most local LIRGs tend to have a disturbed morphology and a high gas fraction \citep{Sand96}. A study done on a small sample of local LIRGs selected from the IRAS Bright Galaxy survey found that all of the LIRGs with $L_{IR}  \geq 10^{11.5}L_\odot$ show some signature of interaction while LIRGs with $10^{11.1}h^{-2}L_\odot  \leq L_{IR} \leq 10^{11.5}h^{-2}L_\odot$ do not show such obvious signature\citep{Ish04}. This work further suggests that the $L_{IR}$ activity of LIRGs appears to track their merger history.
 
 In this paper, we concentrate specifically on the environments of strongly star-forming galaxies, rather than on relations in the global galaxy population. We try to find possible clues for the existence of any fundamental differences between LIRGs  and non-LIRG IR galaxies (IR galaxies with $L_{IR}\leq10^{11}h^{-2}L_\odot$) in the local  Universe. Our results confirm that there is indeed a dramatic density difference above and below  $L_{IR}=10^{11}h^{-2}L_\odot$.  We show that LIRGs specifically behave very differently than the general galaxy population as a function of environmental density, and that this difference is not stellar mass dependent. In  addition, we made a comparison of the  high and low  redshift SFR-density relationships to shed light on the connection between environment  and IR  activity of LIRGs.  Throughout  this work, we used $\Omega_m=0.3$, $\Omega_\Lambda=0.7$ and $H=100h^{-1}$.

 \section{\large Data Samples}

 \subsection{The Point Source Catalogue Redshift (PSCz) Sample}   
 \label{sec:PSCz}

 The Point Source Catalogue redshift survey (PSCz; \citealt{Saun00}) contains 18351 IRAS selected galaxies complete to $S(60)= 0.6Jy$.  The catalogue serves a dual purpose in this work. First, it is used  as an IR target galaxy sample, together with another IRAS catalogue (see section \ref{sec:IIFSCz}), around which environments are measured.  Second, it is used as a tracer of the density field. 

For the first purpose, we cross-match the PSCz catalogue with the 2MASS Extended Source Catalogue \citep{Jarret2000} using a search radius of $60^{''}$ in order to obtain the absolute K-band magnitude $M_K$, which is a good tracer of stellar mass. The choice of the search radius was made based on tests done for different searching angles ranging from $5^{''}$ to $100^{''}$.  The number of galaxies for which there is at least one match increased at a constant rate starting from $5^{''}$ until it reached about $60^{''}$.  We thus adopted $60^{''}$ as a reasonable search radius that can give us genuine matches with minimal inclusion of false matches. Based on the surface density of 2MASS galaxies, we estimate that there is only a $\sim1\%$ chance of having a random 2MASS match with this radius. At $60^{''}$, less than $5\%$ of the PSCz galaxies have two matches and less than $1\%$ have more than two matches; considering the low fraction of multiples, we merely selected the closest one and did not attempt to split the IRAS detection into multiple 2MASS counterparts. After matching the entire PSCz to the 2MASS at this radius and excluding the galactic plane, $|b|\le10$,  and sources  with $V\le1000h$ \kms,  we found  11727 PSCz  galaxies having  matches from 2MASS.  PSCz-I is 80\% complete, missing only 20\% of all PSCz galaxies at $|b|>10$ and $V\le1000h$ \kms. These galaxies form part of the sample of IR-selected galaxies whose environments we study and hereafter we refer to it as PSCz-I.  
We also verified that the non-matched PSCz galaxies follow very similar $L_{IR}$ and redshift distributions as the matched 80\% of the PSCz galaxies, and hence the exclusion is very unlikely to give rise to any biases in the final results.  Out of the total number of PSCz-I, 2956 galaxies  have  $L_{IR}\ge  10^{11}h^{-2}L_\odot$ while  the remaining 8771  have $L_{IR}\le 10^{11}h^{-2}L_\odot$.

Secondly, we use the entire PSCz catalogue, after removing the galactic plane and galaxies with $V~\le~1000$~\kms$\,$ but without matching to 2MASS,  as a tracer of the density field in order to quantify  the local environment around target galaxies.  The final number of galaxies in this case is 14639 and we call this sample as PSCz-II.

 \subsection{The Imperial IRAS-Faint Source Catalogue (FSC) redshift Sample (IIFSCz)} 
\label{sec:IIFSCz}

The IIFSCz catalogue  \citep{Wang09} is constructed based on the IRAS Faint Source Catalogue (FSC) \citep{mosh94} and has 60,303 galaxies selected based on their $60\mu m$ IRAS flux and covers 61\% of the entire sky. Almost 55\% of these galaxies have spectroscopic redshifts found by cross-matching IRAS-FSC  with various previous redshift surveys such as IRAS PSCz \citep{Saun00}, FSSz \citep{oliv96} and 6dF \citep{Jon09}. Approximately 20\% have photometric redshifts estimated either by template-fitting or an empirical training method, but we 
considered only galaxies with spectroscopic redshifts in this work.  At $S(60)\geq0.36Jy$, the catalogue is 90\% complete in redshift.

Since PSCz-I (section \ref{sec:PSCz}) and IIFSCz samples are both originally from IRAS, we think it is natural to merge them together. Before the two samples are merged, 5853 duplicates between PSCz-I and IIFSCz are identified and removed from IIFSCz by cross-matching them with a search radius of $ \leq20^{''}$, which is a reasonable radius based on IRAS positional uncertainties \citep{Neugebauer1984}.  Removing again the galactic disc $(|b|\le 10)$, sources with $V\le1000h$ \kms, and including only sources with a K-band detection and a spectroscopic redshift, the IIFSCz sample is reduced to 16207 galaxies. We refer to this final IIFSCz catalogue as IIFSCz-I. In this subset, 5137 galaxies have $L_{IR}\ge 10^{11}h^{-2}L_\odot$ while the remaining 11069 have $L_{IR}\le 10^{11}h^{-2}L_\odot$. 
Finally PSCz-I and IIFSCz-I are merged and hereafter referred as IRAS-I sample. However, due to incompleteness issues in IIFSCz-I, the merged sample (IRAS-I) is not used as a density tracer but only as an IR target galaxy sample around which the environment is studied. This merged catalogue has 27934 galaxies with a redshift range $0.003<z<0.14$ and a mean redshift of 0.029.
For further details on redshift distributions and flux limits of the samples, readers are advised to consult \citet{Saun00} and \citet{Wang09}, though for reference we note that the flux completeness limit of 0.36 Jy quoted above corresponds to approximately $L_{IR}  = 7 \cdot 10^{9}  L_{\odot}, \  7 \cdot 10^{10}  L_{\odot},$ and $9 \cdot 10^{11} L_{\odot}$ at redshifts $z=0.01$, 0.03 and 0.10, respectively. The corresponding SFR limits calculated using the method below, are 1.3, 12, and 150 $M_{\odot} \:yr^{-1} $, for the three redshifts respectively.

\subsubsection{Calculating $L_{IR}$ and correcting for $F_{IR}$ }

The far-infrared flux $F_{IR}= F_{(8-1000 \mu m)}$ of both PSCz-I and IIFSC-I are estimated based on the 12 $\mu m$, 25 $\mu m$, 60 $\mu m$ and 100 $\mu m$  fluxes using Equation \ref{equ:f_IR} below by \citep{Sand96}.
\begin{equation}
 F_{IR}=1.8\times10^{-14}\{13.48f_{12}+5.16f_{25}+2.58 f_{60}+f_{100}\} [Wm^{-2}]
\label{equ:f_IR}
\end{equation} where $f_{12}$, $f_{25}$, $f_{60}$  and $f_{100}$ are the 12 $\mu m$, 25 $\mu m$, 60 $\mu m$ and 100 $\mu m$ fluxes respectively. We compared the fluxes of galaxies belonging to the original PSCz and IIFSCz (duplicates) and checked if the $F_{IR}$ values are the same. We found that IIFSCz is underestimated by approximately 30\% at the lower flux end $(F_{IR}=10^{-39.5}$ to $10^{-39.8}[Wm^{-2}])$. By comparing the two fluxes we made corrections to the entire IIFSCz fluxes using interpolation. However, the correction above $(F_{IR}=10^{-39.5}[Wm^{-2}])$ is negligible.
 
Finally, the far infrared luminosity, $L_{IR}$, for both PSCz-I and IIFSCz-I is worked out from $L_{IR}=4\pi D_L^2F_{IR} [L_\odot]$ where $D_L$ is the  luminosity distance. 
We checked our results against other calibrations such as \cite{Rieke,Bavouzet} to determine $L_{IR}$ based on 24 $\mu m$ fluxes interpolated from the IRAS measurements. The comparison shows both the systematics and scatter are within $\sim$ 0.25 dex in the value of the derived $L_{IR}$ and do not affect any of our conclusions.

\begin{flushleft}\end{flushleft}
\subsection{6dF Galaxy Survey}
The 6dF Galaxy survey \citep{Jon09} is a southern hemisphere near-infrared-selected galaxy redshift survey of 150000 galaxies. Its initial target list came  from many  smaller surveys including 2MASS. For our purpose, we use it as a tracer of the galaxy density field. This catalogue  does not cover the full sky region of our IR sample, and in this respect it is not as good as our PSCz-II catalogue for estimating environments. However, it has the advantage that it is a much denser sample than PSCz-II and, being K-selected, it may trace environment better than PSCz-II which is IR selected. It has a median redshift of 0.053 and goes as deep as $z\sim0.14$. 

\subsection{The Sloan digital Sky Survey (SDSS) data release 4 (DR4)}
\label{sec:sdss_data}
 This work made use of the SDSS \citep{york00} DR4 \citep{adelman06} in order to compare the SFR-density trends in the local Universe to that of local LIRGs. The SFR for the sample was calculated from the extinction-corrected $H_\alpha$ emission line by \cite{brinchmann04}. The spectroscopic and photometric data were matched using galaxy ID in order to identify galaxies with redshift. In order to avoid any possible aperture effects at relatively small redshifts, galaxies below $z=0.04$ are removed \citep{kewley05} and the final redshift range used is $0.04 \leq z \leq 0.1$ .\\

To summarize, our IR target galaxy sample is a merged sample of the PSCz and the fainter IIFSCz catalogues (IRAS-I). To compute environments, we use both the PSCz (PSCz-II) and 6dF samples. In the case of the SDSS sample, we have used it both as a sample of target galaxies, and as a tracer of the density field. All samples are restricted to regions with $|b|>10^\circ$ and $V>1000$ km $s^{-1}$ and cover similar redshift ranges.

\section{Environment Measures}
\label{sec:environment}
\subsection{Local Density}
\label{sec:Local_Density_measurement}

We use the local density of galaxies as a measure of the environment at any given location.  In doing so, we assume that the galaxy density field traces the mass density field in an approximately monotonic fashion.  In other words, we assume that if the galaxy density is higher in region A than in region B, then the mass density is likely also higher in region A than in region B.
 
 Based on this assumption, we measure the local density around each target galaxy in a cylindrical volume with $2h^{-1}$ Mpc radius and $10h^{-1}$ Mpc length in the redshift direction, where the target galaxy is at the centre of the cylinder. A cylinder to quantify the density of galaxies has been employed in several works \citep[e.g.][]{Kauf04} and has proved to be a reasonable geometry. Our choice of cylindrical geometry is motivated based on two main factors. First, the measurements must be done in a clearly defined geometry with a fixed volume to ensure a consistent and uniform density estimation throughout the sample volume. Second, the method must be robust against redshift space distortion effects, which dominate errors in the density estimates. The radius of the cylinder is chosen in such a way that it is of similar size as that of the immediate density field that affects galaxy star formation. The star formation and other spectroscopic properties of galaxies are strongly related to the size of the host dark matter halo contained in a $\sim1$Mpc radius \citep{Blan06}. However, we used a $2h^{-1}$Mpc radius cylinder because at the distant edge of our samples, the space density of galaxies is very low and a 1Mpc radius would be too small to include statistically sufficient numbers of galaxies. We believe that our choice of $2h^{-1}$ Mpc radius cylinder is both large  enough to accommodate the most important part of the surrounding density field that affects star formation and small enough to exclude too much unnecessary dilution. The length of the cylinder ($10h^{-1}$ Mpc $\sim 1000$ \kms) is such that it is just greater than the typical velocity dispersion ($\sim800h$ \kms) inside clusters of galaxies.

 Densities are estimated from three independent flux-limited samples (SDSS, 6dF and PSCz-II) that each have their own redshift-dependent incompleteness (the galaxy density drops with redshift in flux-limited samples).  Moreover, these samples are subject to edge effects in their sky coverage. In order to compensate for these effects, for each of the three density fields we construct a dense random catalog having the same redshift distribution and sky coverage as the real galaxy sample, but with random RA and Dec coordinates (see \cite{Blanton03, coop05} for detailed discussion of the method used).  To estimate the density around a particular galaxy G using a given galaxy catalogue and its corresponding random catalogue, we count the number of the real galaxies $(N_g)$ and random points $(N_r)$ in a cylinder of $2h^{-1}$ Mpc radius and $10h^{-1}$ Mpc length centered on galaxy G, and take the ratio of the two numbers ($N_g/N_r$).  Since both the random points and the real galaxies have the same sky coverage and redshift distribution, $N_g$ and $N_r$ are both affected by the flux incompleteness and any edge effects in the same way. Therefore, taking their ratio removes both these effects while retaining the inherent clustering nature of the real galaxies. Finally, we normalize this ratio by the value corresponding to the mean galaxy density.  We measure this by counting the number of galaxies ($n_g$) and random points ($n_r$) in cylinders centered on 10000 randomly selected points in space and taking the average of the 10000 values of $n_g/n_r$.  Our final density measurement is thus
\begin{equation}
\rho=\frac{N_g/N_r}{\langle n_g/n_r \rangle}   
\label{eq:density} 
\end{equation}
Estimated in this way, our density estimates are expressed in units of the \textit{mean density} of galaxies and they are insensitive to both redshift and sky incompleteness in the galaxy sample used.

\subsection{Field/group/cluster environment}
\label{sec:Field_Group_Cluster}
In addition to measuring the local density around galaxies, we wish to classify their environments into the traditional field, group, or cluster categories.  We identify galaxy systems in a volume-limited sub-sample of the 6dF data using the \cite{Berlind06} friends-of-friends algorithm.  This algorithm was designed to group together galaxies that live in the same dark matter halo.  We define each system's luminosity to be the sum of the luminosities of all its member galaxies that appear in the volume-limited sample.  We then calculate rough mass estimates for the systems assuming a monotonic relation between their luminosities and the masses of their underlying dark matter halos. By matching the measured space density of 6dF systems to the theoretical space density of dark matter halos (given the concordance cosmological model and a standard halo mass function), we assign a virial halo mass to each system. This procedure ignores the scatter in mass at fixed luminosity and is only meant to yield approximate estimates.  Once we have mass estimates for all the 6dF identified systems, we classify them as field, group, or cluster based on these masses as follows: field galaxies belong to halos with $M_{halo}<10^{13}h^{-1}M_\odot$, group galaxies belong to halos with $10^{13}M_\odot < M_{halo} < 10^{14}M_\odot$, and clusters are halos with $M_{halo}>10^{14}M_\odot$.

\begin{figure}  
\centering \includegraphics[scale=0.23]{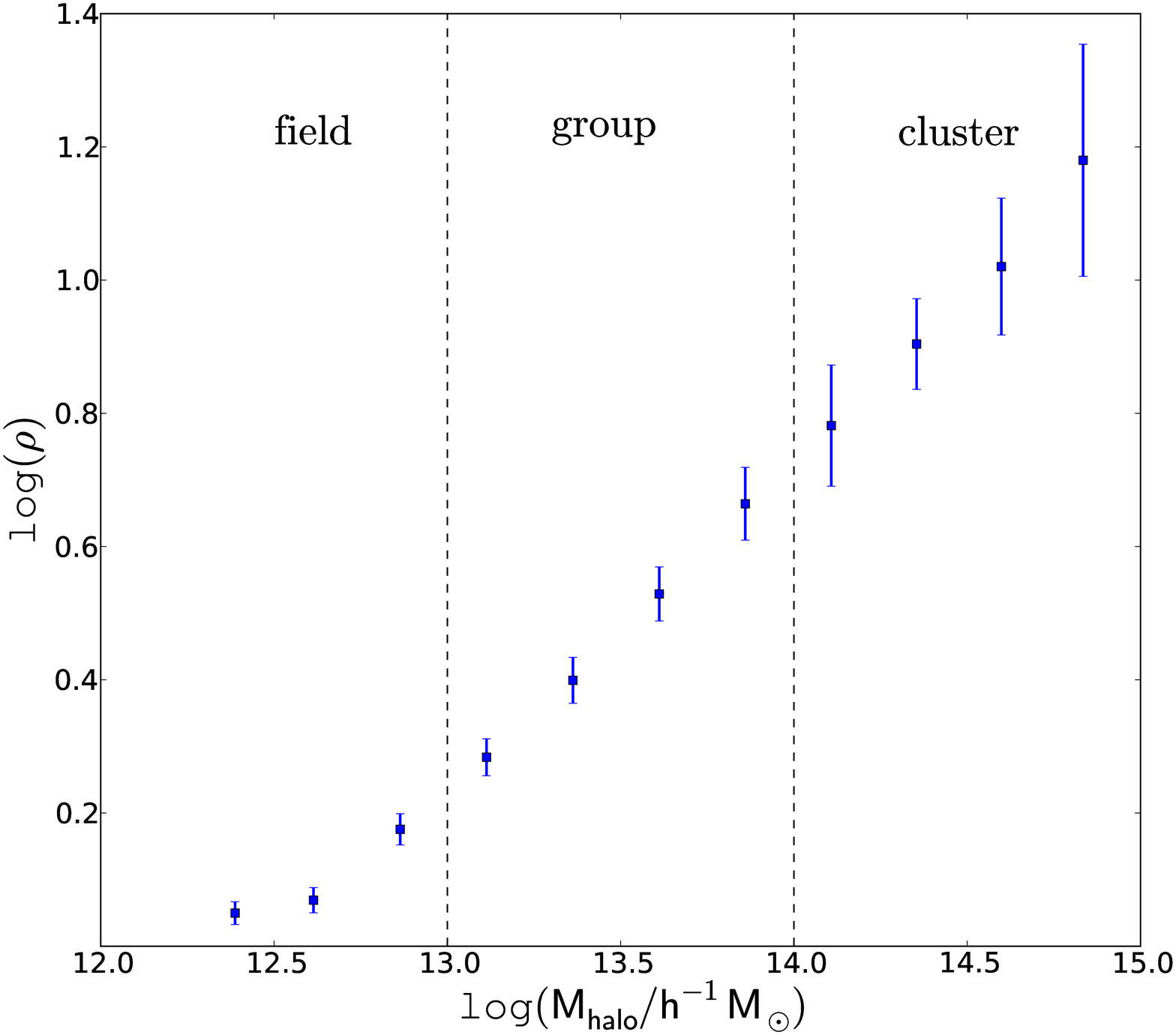}
\caption{The relationship between local density and halo mass for galaxy systems (groups and clusters) identified using a 6dF volume-limited sample of galaxies.  The local density around each system is measured in cylinders of $2h^{-1}$Mpc radius and $10h^{-1}$Mpc length. The error bars represent the uncertainties in the mean estimated from the standard deviation of the density values in each bin and are 4$\sigma$ in their magnitude. The mass of each system is estimated from the abundance of systems with similar total luminosity.  Points show the mean local density in bins of mass.  The vertical dashed lines mark the transitions between the field, group, and cluster regimes, which we place at halo masses of $10^{13}h^{-1}M_\odot$ and $10^{14}h^{-1}M_\odot$.}
\label{fig:group}
\end{figure}

We compute the local density around each system using the full flux-limited 6dF sample, as described in section~\ref{sec:Local_Density_measurement}.  Figure~\ref{fig:group} shows the resulting relationship between local density and halo mass and indicates the two mass scales that separate the field/group and group/cluster regimes.  According to the figure, these mass scales ($10^{13}h^{-1}M_\odot$ and $10^{14}h^{-1}M_\odot$) correspond to local densities of log$(\rho)=0.2$ and log$(\rho)=0.8$.  We have thus determined the 6dF densities that correspond to the field, group, or cluster regimes and we will use them in subsequent results.  Unfortunately, we cannot do the same for PSCz-II because it is too sparse a sample for identifying groups and clusters.

\section{Star formation rate and stellar mass estimates}
\label{sec:SFR_estimation}

The far-infrared luminosity traces the young stellar population of galaxies and can be used in estimating the SFR of IR selected galaxies \citep{Kenn98, Cha01}. This method works very well, especially in the case of late-type highly dust enshrouded star forming galaxies. 

$L_{IR}$ as a SFR indicator is consistent with optical indicators such as $H_\alpha$. \cite{hwang2010} compared the SFR of IRAS selected galaxies using both $L_{IR}$ and $H_\alpha$ (by cross-matching IRAS with SDSS galaxies) and found that the results from the two SF indicators are quite similar. However, at high $L_{IR}$ values, especially in the ULIRG regime, the possibility of AGN contamination \citep{Veil99,nardini09} is an issue of concern.  AGN could lead to overestimation of the true SFR and as a result SFR from $L_{IR}$ and other indicators such as $H_\alpha$ might not agree. 
 
The galaxies used in this work are IR selected (IRAS-I) and we use the method described in \cite{Kenn98} based on the integrated far-infrared light $L_{(8-1000\mu m)}$ to estimate their SFR. The SFR is given as:
\begin{equation}
\mbox{SFR}\;[M_\odot\:yr^{-1}]=L_{IR}/(2.2\times10^{43} \:erg \:s^{-1})=L_{IR}/(5.8\times10^9 \:L_\odot) 
\end{equation}
 
\begin{figure*}  
\centering \includegraphics[scale=0.32]{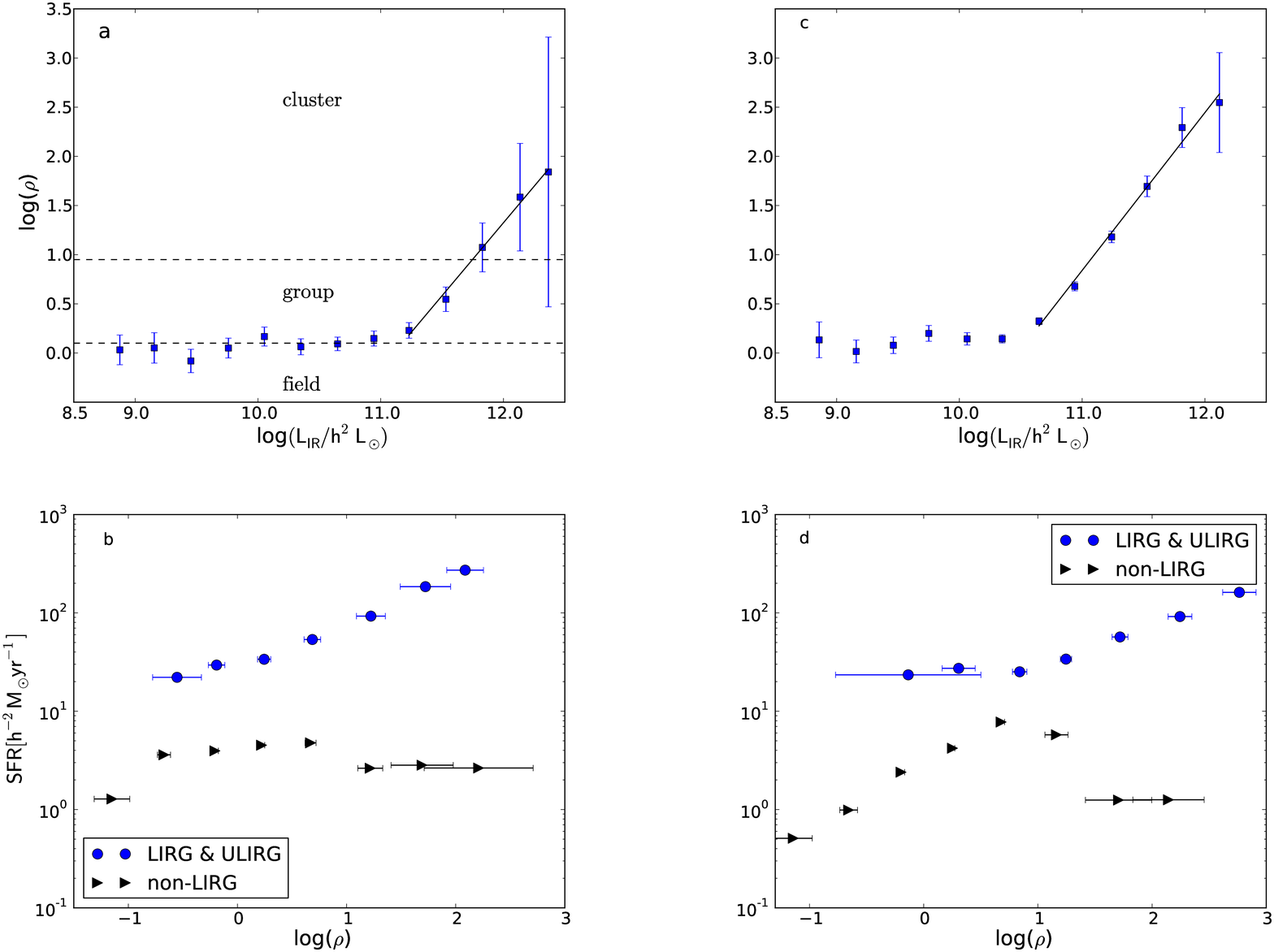}
\caption{{\it Top panels}: The relationship  between $L_{IR}$ and the mean local densities for IRAS galaxies.  The mean local densities are estimated using 6dF and PSCz galaxies in panels (a) and (c), respectively.  The data points show the mean density (as given by Eq.~\ref{eq:density}) in log$(L_{IR})$ bins of width 0.3. The horizontal dashed lines mark the transition from field to group and from group to cluster.  {\it Bottom  panels}: The relationship between SFR and the mean local densities for IRAS galaxies.  The mean local densities are estimated using 6dF and PSCz galaxies in panels (b) and (d), respectively.  The data points show the mean SFR in logarithmic density bins of width 0.5.  Results are shown separately for LIRGs \& ULIRGs ($L_{IR}>10^{11}h^{-2}L_\odot$; circles) and non-LIRGs ($L_{IR}<10^{11}h^{-2}L_\odot$; triangles).  The error bars in the upper and lower panels are 6$\sigma$ and 10$\sigma$ respectively and represent the uncertainty in the mean calculated from the standard deviation of values in each bin.}
\label{fig:sf_LIR_den_panel}
\end{figure*}
 
The K-band light is believed  to originate mainly from old stellar populations and hence it is a good tracer of the stellar mass of galaxies. In this work, we use the \citet{Bell01} relation between K-band absolute magnitude and $M/L$ ratio, which is based on a range of spiral galaxy evolution models, to derive stellar mass. It is worth emphasising that the $M/L$ ratio is not constant but varies with the $M_K$ value of the individual galaxies. We then estimate the value of $M/L$ based on the K-band absolute magnitude of each galaxy, and convert the $M/L$ ratio into stellar mass using $M_{K,\odot}=3.33$. 
 
 At the high K-band absolute magnitude end, the relation gives log$(\frac{M}{L})$ value of  approximately -0.2 but most of the galaxies in this regime are highly star forming galaxies with gas fractions of about $60\%$. According to \cite{Bell01}, a gas fraction of $60\%$ corresponds to a smaller value of $M/L$. Therefore, using log$(\frac{M}{L})=-0.2$ results in a possible overestimation of the mass for the highest star forming galaxies.
 
 In  addition, extinction and contamination from  emission lines (e.g., $Pa_{\alpha}$ and $Br_{\gamma}$) near the K-band could also bias the mass estimation. Typical average K-band extinction in LIRGs is up to $A_K\sim 0.3$ \citep{petri08}, which means that there is a possibility of underestimating the stellar mass by a factor of about $30\%$. The effect from emission lines can be very strong especially with very young stellar populations \citep{zackrisson08} and leads to over-estimation, but it becomes weaker for averaged stellar populations of various age \citep{rhoad98}. 
 As a result, the mass bias from the emission lines and extinction is minimal in most cases and could work in either direction.  
We would like to remind the reader that henceforth when stellar masses related to our sample are discussed, they are derived using the absolute K-band luminosities.

\section{\large Results}
\label{result}
The mean local densities around IRAS-I galaxies are estimated  from  two  separate  redshift samples: 6dF and PSCz-II and the 6dF density estimates are finally classified into: field, group or cluster. The mean local density, SFR and stellar mass estimation are made based on the techniques  described above in Sections~\ref{sec:environment} and \ref{sec:SFR_estimation}. The relationships of these  physical  quantities  with  SF  are explored  with  the  main intention   of  understanding   the connection between SFR  and environments of local LIRGs and non-LIRG IR galaxies. 

We explore how local density and the $L_{IR}$ of local LIRGs and non-LIRG IR galaxies compare. Panels (a) and (c) in Figure~\ref{fig:sf_LIR_den_panel} show the relationship between density and $L_{IR}$ for IRAS-I galaxies with the local density measured using 6dF and PSCz-II galaxies, respectively. In both cases we note  that there is a clear  difference in the environments of LIRGs and non-LIRG IR  galaxies. LIRGs live in an environment which is up to 100  times (panel (a)) and 300 times (panel (c)) as  dense as the environment of non-LIRG IR  galaxies, while non-LIRG IR galaxies live in an environment which corresponds to the mean density (recall that density measurements are made in units of the mean density and within 2Mpc radius). 

\begin{figure}  
\centering \includegraphics[scale=0.3]{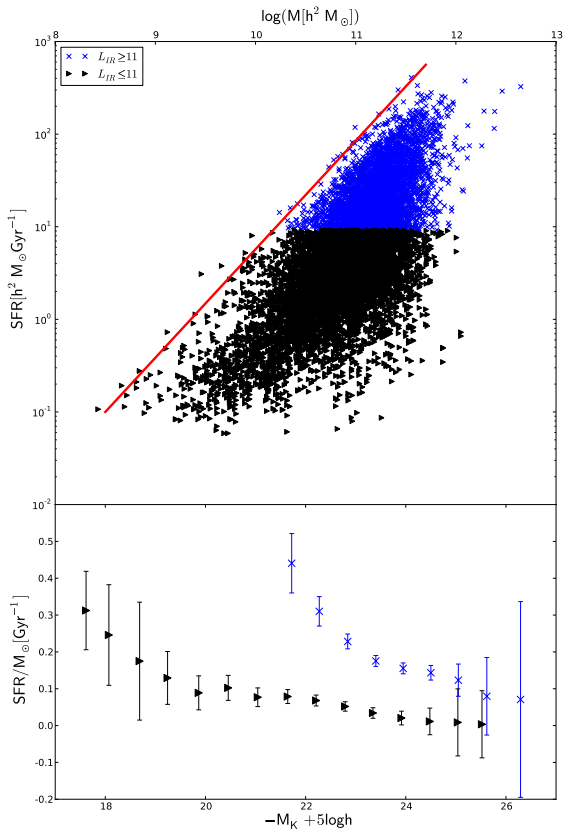}
\caption{{\it Top panel}:  SFR vs. $M_K$ for both LIRGs (blue crosses) and non-LIRG IR galaxies (black triangles). In the upper horizontal axis of the top panel, stellar mass corresponding to the observed $M_K$ values in the lower horizontal axis is presented. SFR correlates with $M_K$ (stellar mass); however, the correlation shows wide scatter especially at the brighter $M_K$ end. The scatter seems to be the result of mixing of two different populations of galaxies (possibly red sequence and blue cloud) having two different relations between SFR and stellar mass. The red diagonal line demonstrates the existence of a physical limit in SFR at a given $M_K$. {\it Bottom panel}: The sSFR (specific SFR) of both LIRGs (blue crosses) and non-LIRG IR (black triangles) galaxies declines slowly with $M_K$.  Moreover, at fixed $M_K$, LIRGs have a higher sSFR than non-LIRG IR galaxies. The error bars in the lower panel are 3$\sigma$ in magnitude and are the uncertainties in the mean value in a given bin calculated from the standard deviation of the specific star formation values.}
\label{fig:ssFR_SFR_M}
\end{figure}

In panel (a) of Figure~\ref{fig:ssFR_SFR_M}, where the density measurements are made from the 6dF sample, we note that the non-LIRG IR galaxies live in an environment between field and group, and the LIRGs mostly live in group environments. The highest $L_{IR}$-LIRGs and the ULIRGs live in cluster environments. The transition from field to group environment is dramatic at $L_{IR} \sim 10^{11}L_\odot$, implying a fundamental connection between environment and this luminosity scale. However, the transition between the environments of LIRGs and ULIRGs is smooth.

Moreover, above $L_{IR} \geq10^{11}h^{-2}L_\odot$ in panel (a) and slightly earlier in panel (c), the density of the LIRG environment increases monotonically with $L_{IR}$. Analytical expressions for the relation between $L_{IR}$ and local density $(\rho)$ when using 6dF and PSCz-II galaxies as a density field are given as:

\begin{equation}
  \mbox{log}(\rho)=1.4\;\mbox{log}(L_{IR}/h^2L_\odot)-15.5
\end{equation} and
\begin{equation} 
\mbox{log}(\rho)= 1.47\; \mbox{log}(L_{IR}/h^2L_\odot)-15.2,
 \end{equation} 
respectively. As it is seen in the equations, using PSCz-II galaxies as a density field produces a slightly steeper relation between $L_{IR}$ and $\rho$ than using 6dF galaxies.

To see the difference between LIRGs and non-LIRG IR galaxies more clearly, we study the relation between SFR and local density for each in the bottom two panels of Figure~\ref{fig:sf_LIR_den_panel}. Panels (b) and (d) represent the splitting of panels (a) and (c) into LIRGs and non-LIRGs IR galaxies and  converting the corresponding $L_{IR}$ values into SFR following the procedure mentioned in section \ref{sec:SFR_estimation}. The LIRGs, according to these plots, exhibit a SFR that increases with density, which is the reverse of what is observed in the local Universe for normal galaxies and similar to the SFR-density relationship at $z\sim1$ found by \cite{Elbaz07} and \cite{coop08}. This trend holds regardless of whether local density is measured using 6dF or PSCz-II galaxies.
 
The upper panel of Figure~\ref{fig:ssFR_SFR_M} shows the correlation between $M_K$ and SFR for both LIRGs (blue points) and non-LIRG IR galaxies (black points).  This suggests a correlation between SFR and stellar mass on average, but that there is wide scatter especially at the high mass end, possibly caused by two populations of galaxies both with high stellar mass but widely different SFRs. Though we have not confirmed this with SDSS colours, these two high stellar mass populations could represent active \textquotedblleft blue" and dead \textquotedblleft red" galaxies.  
In the lower panel, we present the relation between specific star formation rate (sSFR) and $M_K$ for both LIRGs and non-LIRG IR galaxies, strongly suggesting that the sSFR of both populations decline slowly with stellar mass.  In addition, it is clear from the lower panel that an IRAS-selected galaxy is not a LIRG only because it is a more massive galaxy, but due to other physical reasons, whether internal (e.g. gas fraction) or external (e.g. interactions, environmental densities).  Of these possibilities the effect of large scale environment is studied in this paper. Figure~\ref{fig:ssFR_SFR_M} shows that though LIRGs are all massive galaxies, not all massive IRAS galaxies are LIRGs.   Also, the upper envelope (the diagonal line in the figure) characterises a physical limit to SF for a given stellar mass (Kennicutt 1998). 

\begin{figure*}  
\centering
\includegraphics[scale=0.32]{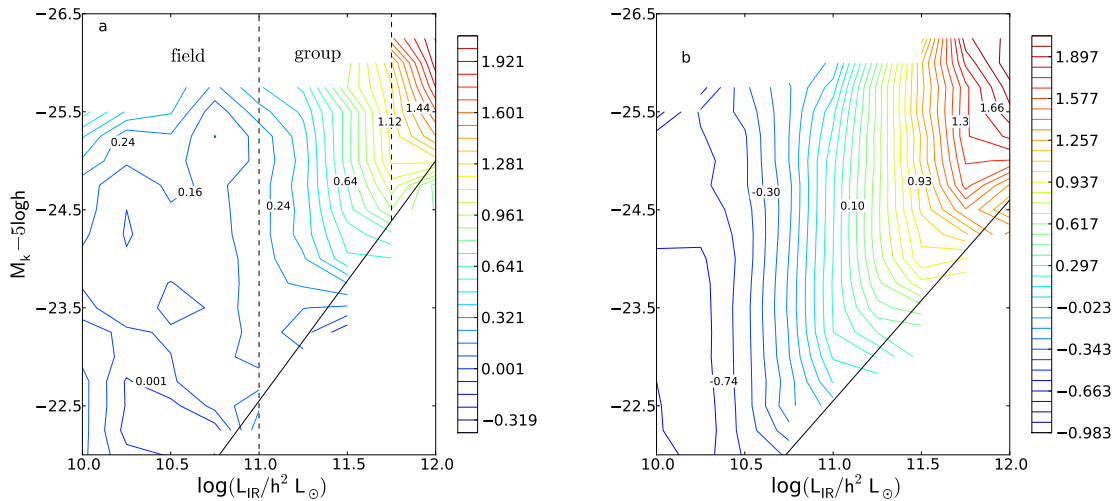}
\caption{Local density as a function of both stellar mass ($M_K$) and SFR ($L_{IR}$).  Local density is measured using 6dF galaxies (panel (a)) and PSCz-II (panel (b)).  Both panels show contours of constant density, with the values of log$\rho$ color-coded to the right of each panel.  The vertical dashed lines in panel (a) denote the approximate transition between field, group, and cluster environments.  In both panels, density contours reflect the mean density in square bins of width 0.25 in both $M_K$ and log$(L_{IR})$.  At constant stellar mass, the SFR of local LIRGs is correlated with their local density. On the other hand, at constant SFR their stellar mass does not show any correlation with local environment.  The solid diagonal lines in both panels highlight that there is a maximum SFR that grows with stellar mass.}
\label{fig: contour}
\end{figure*}

In Figure.~\ref{fig:sf_LIR_den_panel} we showed that there is a connection between $L_{IR}$ (SFR) and the local density around LIRGs, while Figure~\ref{fig:ssFR_SFR_M} showed that there is also a correlation between SFR and $M_K$ for LIRGs.  Since it is generally true that environment correlates with stellar mass \citep{Kauf04}, we now investigate whether the SFR-environment relation for 
the IR-selected galaxies is a product of the SFR-stellar mass or  stellar mass-environment relations, or both.  Figure~\ref{fig: contour} shows the joint dependence of local density on both stellar mass ($M_K$) and SFR ($L_{IR}$).  As before, the two panels show the cases of densities measured using 6dF and PSCz-II galaxies.
 
Above $L_{IR}\sim10^{11}h^{-2}L_\odot$, the density contours show a clear trend in both panels: they are almost perfectly vertical, with each contour corresponding to a specific value of SFR.  This means that SFR depends strongly on local density at fixed absolute K-band luminosity (i.e. stellar mass by derivation), but does not depend at all on stellar mass at fixed density.  It is thus not the case that the correlation between environment and SFR is secondary and caused by a joint dependence on $M_K$, or stellar mass.  Rather, it appears to be stellar mass which does not directly correlate with environment.  Note that by collapsing any $M_K$ bin between approximately -23 and -26 mag into their own density vs.\ $L_{IR}$ plots, each would show the same trend and bend as the relation for the whole sample does in panels (a) and (b) of Figure~\ref{fig:sf_LIR_den_panel}. This is a surprising result and quite different from what is seen with optically and near-IR selected galaxies.

The black diagonal lines in both panels of Figure~\ref{fig: contour} highlight that there is a maximum SFR (or $L_{IR}$) that can be supported in galaxies with a given stellar mass ($M_K$), or conversely, a minimum stellar mass that galaxies must have in order to be IR galaxies (specifically LIRGs).  This is essentially the same limit indicated by the diagonal line in Figure~\ref{fig:ssFR_SFR_M}. We have derived analytical expressions that represent the trend best. The expressions for panel (a) and (b) in Figur.~\ref{fig:ssFR_SFR_M}, respectively, are given as:

\begin{equation}
M_K=-2.45\;\mbox{log}(L_{IR}/h^2 L_\odot)-4.4
\end{equation} and
\begin{equation}
M_K=-1.7\;\mbox{log}(L_{IR}/h^2 L_\odot)-4.75
\end{equation}

\section{\large Discussion}
 
\subsection{Validity of the result}

\subsubsection{PSCz and 6dF as a density fields}
 Since different galaxy types cluster differently, they do not trace the underlying mass distribution of the Universe in the same way \citep{Lahav04}. We have used PSCz and 6dF galaxies to quantify the local density of LIRG and non-LIRG IR galaxies. It has been shown that PSCz (IRAS) galaxies and mass are not very strongly correlated \citep{Taylor01}. On the other hand, 6dF galaxies being near-infrared selected, are likely more strongly biased. It is thus very important to be careful when considering the results found using PSCz-II samples as a density field.
 
Nevertheless, despite the difference between 6dF and PSCz samples as overdensity  tracers, there appears to be no serious discrepancy between the results found using the two density fields in this work. The upturn point in the $L_{IR}$-density relationship occurs at slightly above $10^{11}L_{\odot}$ when 6dF galaxies are used as an overdensity tracer, but when PSCz galaxies are used, it occurs slightly below $10^{11}L_{\odot}$. In the contour plots, when PSCz is used as overdensity tracer we see a strong well-defined pattern but 6dF produces a slightly weaker pattern. Overall, given the main intention of this work which is to find any possible trends using both PSCz and 6dF galaxies as a density tracer, the two samples produced consistent results.

\subsubsection{Testing our density estimation method}

\begin{figure}  
\centering \includegraphics[scale=0.23]{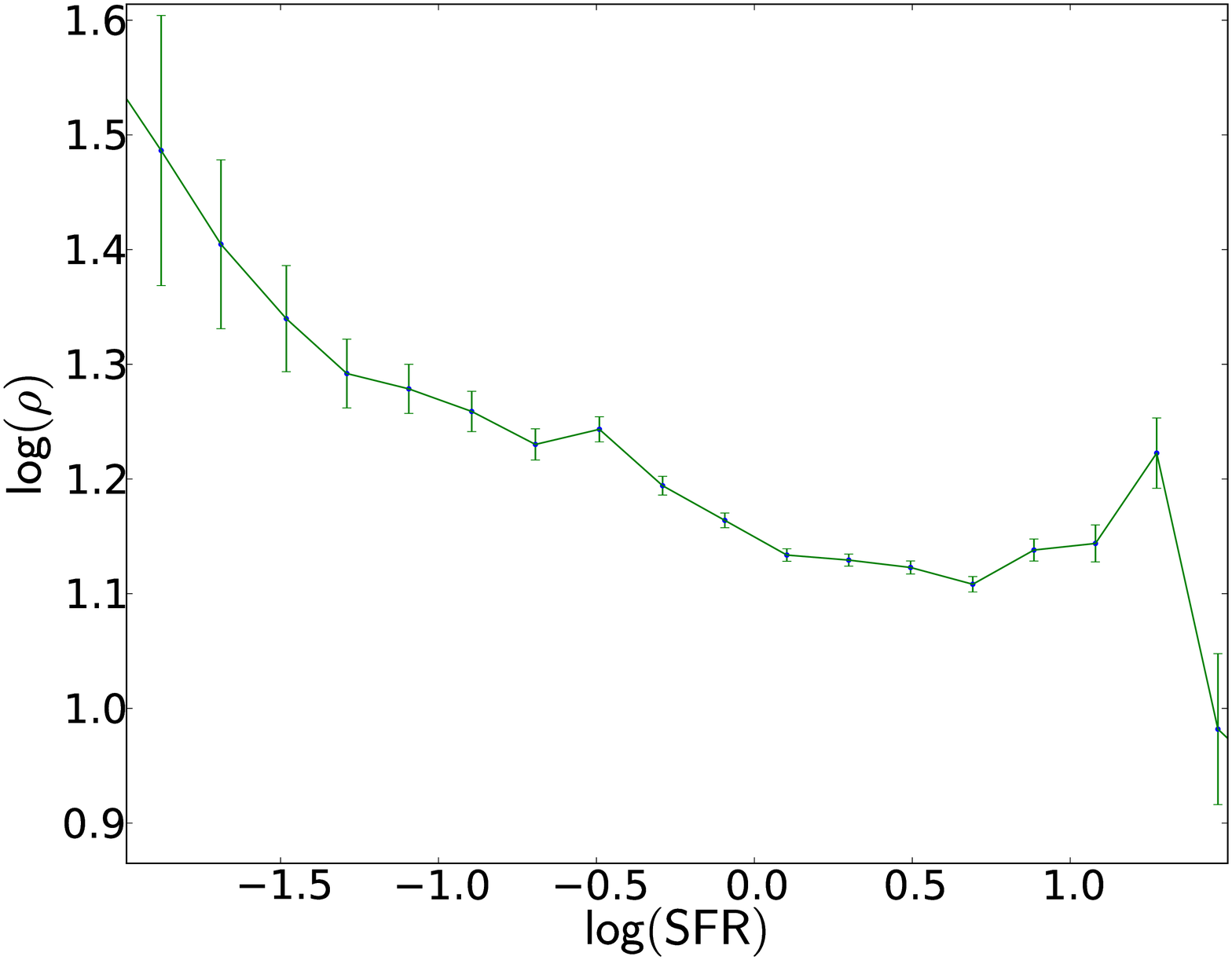}
\caption{Local density as a function of SFR for SDSS galaxies.  The curve shows the mean local density in log(SFR) bins of width 0.2 and error bars show the uncertainty in the mean.  We use SDSS galaxies themselves to estimate local density in units of the mean density of SDSS galaxies.  The relation reflects the normal average SFR-environment relationship found in the local Universe. The SFR is in units of $M_\odot yr^{-1}$. The error bars represent the uncertainty in the mean as calculated from the standard deviation from the density values in each bin.} 
\label{fig:sfr_density_sdss}
\end{figure}

We tested our density estimation method using a flux-limited SDSS sample to check if the method reproduces the expected trend. Following the procedure outlined in section~\ref{sec:Local_Density_measurement}, we estimated the density around SDSS galaxies in cylinders of $2h^{-1}$Mpc radius and $10h^{-1}$Mpc length using SDSS galaxies themselves as the density field. The density measurements in this case are thus in units of the mean density of the SDSS galaxies.  The estimated density versus SFR relation is shown in Figure~\ref{fig:sfr_density_sdss} and according to this result, the SFR decreases monotonically with density. This result is in agreement with other results done on SDSS galaxies and it is also in agreement with what is normally expected in the local Universe \citep{gomez03, Kauf04}.

We also tested whether the size of the cylinder used for the density estimation has an effect on the results. In order to check this, the $L_{IR}$-density relationship in panel (a) of Figure~\ref{fig:sf_LIR_den_panel} was reproduced using cylindrical geometries of different radii: 1, 2.5, 3 and 4 Mpc. All the cases show the same trend, though the 1 Mpc radius result seems to suffer from lower statistics especially at higher $L_{IR}$, and the 3 and 4 Mpc radii results show a somewhat weaker trend, perhaps because of dilution effects.

Our density calculation so far is based on the local number count (over-density) of galaxies and does not take stellar mass effects into account. It is therefore important to check whether the result found above is sensitive to stellar mass weighted density measure. We estimated the over-density of IRAS-I galaxies using the 6dF sample following the same procedure presented in Section \ref{sec:environment}, but this time weighting the contribution of individual galaxies to the density estimate by their $M_K$ value, i.e. essentially their mass. The stellar masses of the random point galaxies used to normalise the densities and deal with the incompleteness were randomly selected from identical redshift and magnitude distributions as the real galaxies. The trend found between $L_{IR}$ and density, as presented in Figure~\ref{fig:sf_LIR_den_panel}, remains the same with the same break point, and hence our results do not change when weighing the density with absolute K-band magnitude.

\subsection{The Relevance of $L_{IR}=10^{11}L_{\odot}$}

We have shown that local LIRGs live in a denser environment than non-LIRG IR galaxies. According to the upper two panels of Figure~\ref{fig:sf_LIR_den_panel}, the far-IR luminosity, $L_{IR}=10^{11}h^{-2}L_\odot$, is not only a matter of numerical convenience to distinguish  between the  two populations of IR galaxies. This luminosity also marks a clear transition in the environments of IR selected galaxies. This possibly implies that LIRGs are a different galaxy population than other IR galaxies with $L_{IR}\leq10^{11}h^{-2}L_\odot$ and the difference is at least partly environmentally triggered.

 Several previous works observed a morphological difference between moderate and high IR-emitting galaxies. However, they were unable to determine where the exact boundary is.  Almost all E's and S0s exhibit insignificant amount of IR activity with $L_{IR}\leq10^9L_{\odot}$ \citep{dejon84}, whereas spirals produce $L_{IR}\geq10^8L_{\odot}$ \citep{Riek86}. \cite{soif84} found that up to 25\% of LIRGs are peculiar or interacting while IRAS galaxies with $L_{IR}>3\times10^{12}L_{\odot}$ are entirely classified as interacting or merging \citep{Sand96}. Similarly, \cite{Menl90} showed that the fraction of interacting systems increases from nearly 10\% to 100\% for $L_{IR}$ value increase from $10^{10.5}-10^{11}$ to $10^{12}$ and above. Putting together sequentially the $L_{IR}$ values of these galaxies with their corresponding morphology, there is indeed indication of a tie between their IR activity and degree of interaction. 

If the level of morphological asymmetry/interaction of galaxies reflects the environment they live in \citep[see e.g.][]{ellison2010}, then the morphological transition luminosity $L_{IR}=10^{11}L_{\odot}$ should also mark a transition in the environments of IR-selected galaxies. Our results confirm that there is indeed a dramatic density difference above and below $L_{IR}=10^{11}L_{\odot}$. 

From this result, it is clear that there is a certain environmental threshold above which galaxies are expected to experience the  extreme SFR seen in LIRGs, though it may depend on other physical characteristics as well (for example, gas fraction and/or the mode of star-formation, see e.g. \citet{caputi09b}). Why exactly $L_{IR}=10^{11}L_{\odot}$? This question is beyond the scope of this work, and needs detailed studies of the physical characteristics of populations of galaxies above and below the $L_{IR}=10^{11}L_{\odot}$ luminosity limit and the local mass distribution around them. 

\subsection{Correlation between SFR and environments of local LIRGs}

Several photometric studies of the morphology of local LIRGs agree that local LIRGs have a highly disturbed  morphology, which is a signature of interactions they have undergone \citep{Sand96}. However, it was not clear in what fashion the environments and the IR activity of local LIRGs  are related. We found that their environment is not only denser than that of non-LIRG IR galaxies, but we also found that $L_{IR}$ (SFR) of LIRGs is a monotonically increasing function of the richness of their local environment (Fig.~\ref{fig:sf_LIR_den_panel}).
 
An independent work by \cite{Menl90} found that the morphology (interaction stage) of (U)LIRG systems is a monotonic function of $L_{IR}$. Since merger or interaction probability is a function of environmental richness (though not in a simple monotonic way), the morphology-IR activity witnessed in (U)LIRGs is the direct reflection of the IR-environment correlation shown in this work. Therefore, the IR  activity  of LIRGs is at  least partly interaction driven and also their $L_{IR}$ (or SFR) can be considered as a tracer of their interaction stage. The same trend has been witnessed among $z\sim1$ LIRG groups. \cite{Elbaz07} found in the GOODS field that the intrinsic luminosity of LIRGs increases with the local galaxy density, so that the most luminous ones live in the center of a cluster that is in the process of formation and where interactions are strongest.

Contrary to our finding, \cite{hwang2010} using IR galaxies with $L_{IR}\geq 10^{10}L_\odot$ found that  SFR is not correlated with environment and concluded that the SFR-density relationship of these galaxies are similar with local galaxies.  However, the results need not necessarily be in contradiction, since the presence of galaxies with $L_{IR}\leq 10^{11}L_\odot$ in the $L_{IR}$ vs.\ density plot in \citet{hwang2010} would weaken the relationship we have observed {\it specifically} for LIRGs  at  $L_{IR}\geq 10^{11}L_\odot$ in Figure~\ref{fig:sf_LIR_den_panel}.  Also, the density measurement they have used is quite different with the one we have used. 

\subsection{Correlations between stellar mass, SFR, and environment} 

According to Figure~\ref{fig: contour}, LIRGs exhibit a correlation between their $L_{IR}$ (measure of SFR) and environment at constant $M_K$ (measure of stellar mass) but they do not show any clear correlation between $M_K$ and density at constant $L_{IR}$. No matter what the K-band absolute magnitude (or stellar mass) is, local LIRGs in similar density environment tend to have the same SFR, as defined by their IR luminosity. The implication is that the SF property of local LIRGs is controlled by their environment regardless of their stellar mass.  This is exactly the opposite to what is observed with the global galaxy population in the local Universe, where it has been shown that it is rather the {\it stellar mass} which is driving the apparent SFR vs.\ density relations, seen e.g.\ in a similar contour plot in \citet{peng10} (their Fig.~2).  Such differing results compared to the general galaxy population should not necessarily be surprising since our sample is selected purely on their IR-properties, i.e.\ their SF properties.  Nevertheless, it is interesting to examine further where the differences might come from. Is there, for example, some implicit mass-selection effect working which would explain the results?

\begin{figure}  
\centering \includegraphics[scale=0.33]{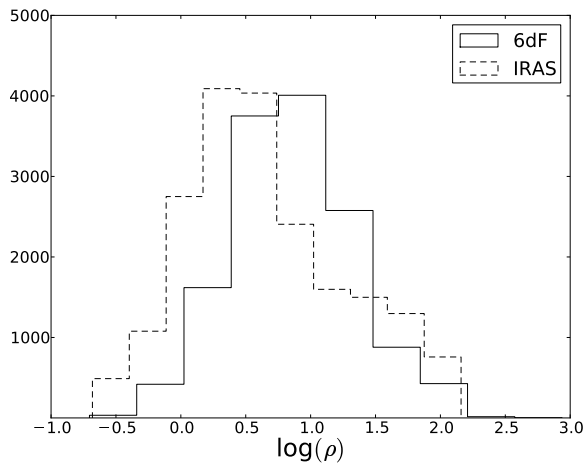}
\caption{The density distribution of 6dF and IRAS galaxies estimated with 2 Mpc radius and 10 Mpc length cylinder from 6dF density field.  Both samples have $M_K$ values in between -24.5 and -25.8 mag. The histogram of the IRAS sample is scaled six times for an easier comparison.}
\label{fig: massdistribution}
 \end{figure}

We compared the density distribution of IRAS-I sample galaxies with those from the 6dF {\it in identical mass ranges}. As a K-selected sample, the 6dF galaxies are assumed to be representative of the general galaxy population. In both cases the densities are estimated from 6dF sample.  As an example, Figure~\ref{fig: massdistribution} presents the distribution of the densities around 6dF galaxies and IRAS galaxies calculated exactly the same way as the previous density measurements in the same cylinder sizes, but now for a restricted $M_K$ range of -24.5 to -25.8 mag for both samples.  The distributions largely overlap.

Firstly, the direction of the relatively small density difference between the distributions seen in Figure~\ref{fig: massdistribution} is understandable considering the parent populations. The IRAS sample consists of preferentially star-forming galaxies, which are known to live, in general, in lower densities than an average K-selected population. That the very highest densities are missing in the IRAS sample must mean that extremely high SFRs are not found in the local universe cluster cores. 

More interesting, however, is the overall similarity of the IRAS and 6dF density distributions and what it means.  At this fixed mass bin the general galaxy population consists mostly of passive red sequence galaxies regardless of their environmental density \citep{Kauf04}. On the other hand, the IRAS galaxies in this mass and density range are typically strong LIRGs with SFR $\sim100 \ M_{\odot} yr^{-1}$ (see Figs.~\ref{fig:sf_LIR_den_panel} and~\ref{fig: contour}).  Hence,  the wildly different SF characteristics of these two samples, presumably due to very different gas contents,  is not strongly related to either stellar mass or environment.   Meanwhile, however,  we had already shown that the SF characteristics of the IR-selected galaxies are {\it driven by environment and not by stellar mass} (Fig.~\ref{fig: contour}) and that the SF increases with density (Fig.~\ref{fig:sf_LIR_den_panel}).  This result is clearly in contrast to K-selected galaxies in general, whose SF properties are driven by stellar mass \citep[e.g.][]{peng10}, and whose SF declines with density (e.g.\ Fig.~\ref{fig:sfr_density_sdss}).  LIRGs behave differently to normal K-selected galaxies as a function of environment, and this contrasting behavior is not dependent on some hidden stellar mass-bin or environmental selection, as  shown by Figure~\ref{fig: massdistribution}.

\subsection{The reversal of SFR-density trend in the local Universe and the possible origin of local LIRGs}
\label{sec:reversal} 

The SFR-density relationship found for SDSS galaxies ($z\sim0$ galaxies) in Figure~\ref{fig:sfr_density_sdss} is  consistent with results found for SDSS galaxies by other works and it is also the reverse of what is observed for blue cloud galaxies at $z\sim1$ \citep{coop08,Elbaz07}. At $z\sim1$, SFR decreases with environment for red sequence galaxies and then increases for blue sequence galaxies (V-shaped trend, \citet{coop08};  note, however, uncertainties in this result with different mass bin selections \citep[e.g.][]{patel09,sob2010}). The highest SF (bright massive blue galaxies) and lowest SF (massive red sequence galaxies) are both found in denser environments than the intermediate SF galaxies. 

It is believed that the highest SF population has possibly evolved into the red-sequence galaxies at $z\sim0$ and resulted in the major difference of SFR-density relationship at the two redshifts, the absence of a population of galaxies whose SFR increases with density.

The SFR-density trend we observed for local LIRGs is similar to the relationship observed for $z\sim1$ blue galaxies: SFR increases with local density (see Fig.~\ref{fig:sf_LIR_den_panel}). Combining the decreasing SFR-density relation found for SDSS galaxies (see Fig.~\ref{fig:sfr_density_sdss}) and the increasing result for local LIRGs (see Fig.~\ref{fig:sf_LIR_den_panel}), we construct a complete picture of SFR-density relationship at the local Universe.
 
This trend is identical with what is observed at $z\sim1$ according to \citet{coop08}: the SFR first decreases and then increases with local density. The only and crucial difference between the two SFR-density relationships is that at the highest SF end of the relationship there are much fewer galaxies locally than at $z\sim1$. 

The similarity of the SFR-density trend between LIRGs and blue cloud galaxies at $z\sim1$ implies a possible similarity of their formation process. This together with the sharp drop of the number density of LIRGs since $z\sim1$ could mean either that local LIRGs might be survivors of whatever process transformed the blue population of galaxies at $z\sim1$ to the present, or they are formed with the same physical process as $z\sim1$ blue galaxies but only at later time.

The positive correlation of SFR and density for LIRGs raises questions regarding the mechanism of intense bursts of star formation in the local Universe. It is commonly accepted that galaxy-galaxy interactions cause starbursts, but not larger scale environments studied in this work. Normally, as density increases, the possibility of finding processes such as ram-pressure stripping, galaxy harassment and strangulation that suppress SF is high. Despite this expectation, we see enhanced SF as density increases around local LIRGs and ULIRGs on 2 Mpc scales. It is even appears that higher $L_{IR}$ LIRGs live in an environment near the transition between group and cluster and ULIRGs are found in cluster environment. The result we have found is observationally supported with the presence of LIRGs in environments such as the outskirts of clusters \citep{Laag06, duc04, lemon98, Kleinmann88} that are not normal for the magnitude of SFR they exhibit. Note however that with our adopted density scale of 2 Mpc we do not isolate for example cluster cores; this will be addressed in a future work.
 
A simulation study by \cite{Martig08} found that major mergers or interactions happening near the tidal fields of cosmological structures such as groups and clusters are more efficient in producing the highest levels of star formation. According to their work, two M33-like galaxies interacting in the tidal field of Local-group-like or Virgo-like cluster could have their star burst raised by a factor of two or more. Given the environment where LIRGs and ULIRGs are found in our study, star formation mechanisms such as tidal field-induced SF by cosmological structures could explain their star formation mechanism. 

\section{\large Summary and Conclusion}

 We have studied the connection between the star formation and the environments of local LIRGs in comparison to other types of local and high redshift galaxies in search of understanding regarding the nature of their SF-environment relation and also their possible origin and evolution. Both the LIRG and non-LIRG IR galaxies are drawn from IRAS catalogues. PSCz and 6dF samples are used to quantify the density fields around the target galaxies in a cylinder of $2h^{-1}$ Mpc radius and $10h^{-1}$ Mpc length. We have arrived at the following main conclusions:

\begin{enumerate}

\item  The IR luminosity value $L_{IR}\sim10^{11}h^{-2}L_\odot$ is a special transition point between sharply different environments among IR galaxies.  We show that there is a dramatic density rise above $L_{IR}=10^{11}h^{-2}L_\odot$ (LIRGs) while the IR galaxies below this luminosity point have a local density equal to the mean galaxy density of the universe.

\item We find that the $L_{IR}$ values of local LIRGs are correlated with their density.  Even if we could not cover the entire ULIRG luminosity range because of lack of data, we noticed that the relation between $L_{IR}$ and density goes in the same manner past the $10^{12}h^{-2}L_\odot$ limit and extends to $10^{12.5}h^{-2}L_\odot$.  We also note that local non-LIRG IR galaxies, LIRGs and low $L_{IR}$ ULIRGs live in fields (or near the transition between group and field environments), group, and cluster environments respectively.  However, it is not clear where exactly within their respective environment (at the centre or outskirts) they are living. This will be addressed in another paper (Tekola et al., in prep.).

\item We find that at constant stellar mass, which is derived from absolute K-band magnitude $M_K$, SFR,  as derived from $L_{IR}$,  is correlated with background density for LIRGs. For non-LIRG IR galaxies this trend is not obvious. On the other hand, at constant $L_{IR}$ value, no trend is observed between $M_K$ and density. The rising trend of SFR of LIRGs with density is in contrast with galaxies in general, and does not depend on mass-bin selection.  In addition, it is found that IR-galaxies seem to require a certain minimum stellar mass at a given SFR  for their existence and this stellar mass limit is linearly correlated with $L_{IR}$. Once the minimum stellar mass limit is met and they become LIRGs, any additional mass they have does not have any effect on their star formation rate.  To form stars at a higher rate, their environment must be elevated. 

\item The local mass overdensity within 2 Mpc scale plays a crucial role in the SFR of local LIRGs.  It thus appears that it is not only the immediate surroundings (one-to-one interactions and merging) of LIRGs that make them LIRGs.  

\item Below $L_{IR}\leq 10^{11}h^{-2}L_\odot$, non-LIRG IR galaxies do not only live in less dense environments than LIRGs, but they experience a flat $L_{IR}$-density trend in 2Mpc scale. Once the LIRG luminosity limit is passed in the lower direction, the 2Mpc scale potential well has no effect on the SFR of IR galaxies in that regime.

\item The SFR of LIRGs is proportional to their environment. This relationship between SF and environment of LIRGs is similar to what is suggested in some studies for blue cloud galaxies at $z\sim1$. This similarity suggests that the formation mechanism of local LIRGs and $z\sim1$ blue galaxies might be similar. The two populations may have formed via similar processes but at different times. Alternatively, local LIRGs might be survivors of whatever process transformed $z\sim1$ blue cloud galaxies into the present red sequence galaxies.

\end{enumerate}
\textit{Acknowledgement} A.G.T. acknowledges Cameron McBride, Russell Johnston and Peter Johansson for their valuable discussions and suggestion. He wants also to extend a special thank to the entire Vanderbilt University astronomy group especially the bridge community and the Ex-gal research group for giving him warm hospitality during his one year visit in which part of this work was carried out.

\bibliography{paper}

\end{document}